\documentclass[titlepage,twocolumn,superscriptaddress,nofootinbib]{revtex4-1}
\usepackage{amsmath}
\usepackage{amssymb}
\usepackage{graphicx}
\usepackage{color}
\usepackage{subfigure}
\usepackage{verbatim}
\usepackage{dcolumn}
\usepackage{bm}
\usepackage{epsf}
\usepackage[colorlinks=true,citecolor=blue,linkcolor=blue,urlcolor=blue]{hyperref}
\usepackage{lmodern}
\usepackage{tikz}
\def\be{\begin{equation}}
\def\ee{\end{equation}}
\def\bea{\begin{eqnarray}}          
\def\eea{\end{eqnarray}}
\def\bi{\begin{itemize}}
\def\ei{\end{itemize}}

\newcommand{\jd}[1]{{\color{red}{{#1}}}}

\newcommand{\tad}{\tau_{\rm AD}}
\begin{document}
\title{    
Defects in Quantum Computers
}
\author{Bart{\l}omiej Gardas}
\affiliation{Theoretical Division, LANL, Los Alamos, New Mexico 87545, USA}             
\affiliation{Institute of Physics, University of Silesia, 40-007 Katowice, Poland}     
\affiliation{Instytut Fizyki Uniwersytetu Jagiello\'nskiego,
             ul. {\L}ojasiewicza 11, PL-30-348 Krak\'ow, Poland}        
\author{Jacek Dziarmaga} 
\affiliation{Instytut Fizyki Uniwersytetu Jagiello\'nskiego,
             ul. {\L}ojasiewicza 11, PL-30-348 Krak\'ow, Poland}
\author{Wojciech H. Zurek}
\affiliation{Theoretical Division, LANL, Los Alamos, New Mexico 87545, USA}    
\author{Michael Zwolak}
\affiliation{Theoretical Division, LANL, Los Alamos, New Mexico 87545, USA}    
\date{\today}
\maketitle

{\bf 
The shift of interest from general purpose quantum computers to adiabatic quantum computing or quantum annealing calls for a broadly applicable and easy to implement test to assess how quantum or adiabatic is a specific hardware. Here we propose such a test based on an exactly solvable many body system -- the quantum Ising chain in transverse field -- and implement it on the D-Wave machine. An ideal adiabatic quench of the quantum Ising chain should lead to an ordered broken symmetry ground state with all spins aligned in the same direction. An actual quench can be imperfect due to decoherence, noise, flaws in the implemented Hamiltonian, or simply too fast to be adiabatic. Imperfections result in topological defects:  Spins change orientation, kinks punctuating ordered sections of the chain. The number of such defects quantifies the extent by which the quantum computer misses the ground state, and is, therefore, imperfect.}


\section{Introduction}

Adiabatic quantum computing~\cite{Nishimori98,Farhi00,Aharonovetal} -- an alternative to the quantum Turing machine paradigm -- is at its core very simple and very quantum: Evolve a system from the ground state of an ``easy'' Hamiltonian $H_0$ to the ground state of $H_1$ that encodes the solution to the problem of interest by varying the parameter $s$ from $0$ to $1$ in 
\be 
H(s)=(1-s)H_0+sH_1.
\label{Hs}
\ee
When $H(s)$ varies slowly enough the system will remain in its ground state, and the answer can be ``read off'' through a suitable measurement of the final state. 

It has always been appreciated that adiabatic quantum computing will be difficult. For instance, even if the hardware to accurately implement $H(s)$ and measure the final (likely, globally entangled) state were available, how slow is ``slow enough'' to retain the system in the ground state? This is a difficult question, as $H(s)$ is likely to have -- somewhere between $H_0$ to $H_1$ -- a narrow energy gap $\Delta$ analogous to the critical point of a quantum phase transition in a finite system. The exact size and properties of such a gap are {\it ab initio} unknown. Yet, for the computation to succeed, this gap should be traversed slowly, on a timescale longer than $\hbar/\Delta$.

Here, we put forth a simple test based on the behavior of the exactly solvable quantum Ising chain in transverse field and deploy it on the D-Wave chip. As we shall see, in addition to the issues of adiabaticity and accessibility of global ground states, there are other practical considerations that affect performance of D-Wave computers, and are likely to play a role in similar devices. 

There are several efforts that aim at such hardware~\cite{Ladd10}. The D-Wave computer is already available and is the obvious guinea pig that we can test. There are by now several papers that, with varying degrees of success, model the behavior of D-Wave~\cite{Lanting14}. We applaud such efforts, but aim at a rather different goal -- a general TAC. 

\begin{figure}[b!]
\includegraphics[width=\columnwidth]{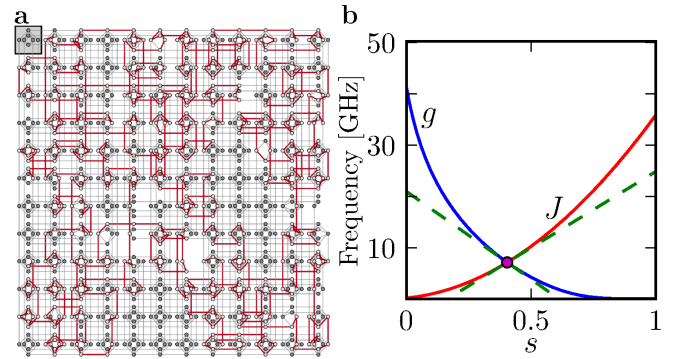}
\caption{
{\bf The quantum Ising chain implemented in a D-Wave computer.}
{\bf a.} 
An example of an Ising chain on the D-Wave ``chimera graph''. The red lines are active couplings between ``spins''. 
{\bf b.} 
A typical annealing protocol for a D-Wave annealer. Here $J(s)=J_{\text{max}}\cdot j(s)$, where
$j(s)$ is a predetermined function increasing from $j(0)=0$ to its maximal value $j(1)$ and
$J_{\text{max}}\in[-1,1]$ is a free parameter that can be turned at will.
}
\label{fig:chimera}
\end{figure}

The quantum Ising chain has a Hamiltonian, 
\be 
H=-g(t)\sum_{i=1}^{L}\sigma^x_i-J(t)\sum_{i=1}^{L-1}\sigma^z_i\sigma^z_{i+1},
\label{Hising}
\ee
in the form of Eq.~\eqref{Hs}. It can be implemented on the D-Wave computer, see Fig.~\ref{fig:chimera}. When the transverse field $g$ and coupling $J$ vary, the ground state of the quantum Ising chain can undergo a transition from a non-degenerate, paramagnetic state, $| \cdots \rightarrow\rightarrow\rightarrow \cdots \rangle$, to a degenerate, ferromagnetic state spanned by $| \cdots \uparrow\uparrow\uparrow \cdots \rangle$ and $| \cdots \downarrow\downarrow\downarrow \cdots \rangle$. The phase transition occurs when $g=J$. The ground state on the broken symmetry side is a defect-free, ferromagnetically ordered ground state. Quenches that are too fast to be adiabatic, or are in some other way imperfect, would instead lead to a ``defective'' state with ``kinks'', e.g., $| \cdots \uparrow\uparrow\downarrow\downarrow \cdots \rangle$.

\begin{figure}[tp!]
\includegraphics[width=\columnwidth,clip=true]{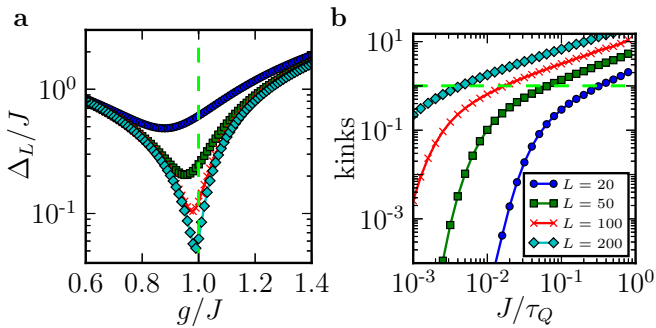}
\caption{
{\bf Adiabaticity in the quantum Ising chain}
{\bf a.}
the relative energy gap $\Delta_L/J$ in Eq.~(\ref{DeltaN}) as a function of the relative transverse field $g/J$. For long chains the gap has a minimum when $g/J=1$.
{\bf b.}
The number of kinks in a chain of length $L$ after a quench with a quench time $\tau_Q$. The dependence crosses over from the power law, Eq.~(\ref{onehalf}), to the Landau-Zener formula, Eq.~(\ref{LZ}), at $\tad$ in Eq.~(\ref{tauQAD}).
}
\label{fig:LZtoKZM}
\end{figure}

\section{Test of adiabatic computing}
The dynamics of quantum phase transitions was first understood by analyzing the density of kinks in the final post-transition state ($g=0$, $J$ at its maximum) as a function of the quench timescale $\tau_Q$~\cite{Dorner2005,Dziarmaga2005,AdvPhys}. Near the critical point, $g=J \equiv J_c$ (see Fig.~\ref{fig:chimera}b and Fig.~\ref{fig:LZtoKZM}a), quench is well approximeted by 
\be 
\frac{g(t)}{J(t)}-1 \approx \frac{t}{\tau_Q}.
\label{ramp}
\ee
That analysis dealt with the limit of very long chains ($L\gg1$)~\cite{Dorner2005,Dziarmaga2005,AdvPhys} where the generation of kinks was a foregone conclusion. However, we are interested in relatively short chains where there is a chance for adiabaticity to survive. This is determined by the gap size, $\Delta_L$, see Fig.~\ref{fig:LZtoKZM}a. At the critical point, $s=s_c$, where
\be 
\Delta_L = J_c\frac{2\pi}{L},
\label{DeltaN}
\ee
the ground and first excited states (that can accommodate a single pair of kinks) undergo an anti-crossing, where the probability of exciting a pair of kinks is given by the Landau-Zener (LZ) formula~\cite{Dorner2005,Dziarmaga2005}
\be 
p=\exp\left(-2\pi^3 J_c\tau_Q/ \hbar L^2\right).
\label{LZ}
\ee
Thus, when $\tau_Q$ exceeds
\be 
\tad=\frac{\hbar L^2}{2\pi^3 J_c},
\label{tauQAD}
\ee
we expect exponential suppression of kinks, i.e., quantum annealing should lead to the ``correct answer'' (in this case, all spins pointing in the same direction). 

When the condition for adiabaticity is not met, $\tau_Q\ll\tad$, the quench timescale also governs the density of excitations according to 
\be 
\frac{1}{2\pi}\frac{1}{\sqrt{2J_c\tau_Q/\hbar }}
\label{onehalf}
\ee
for sufficiently long closed chains~\cite{AdvPhys}. The scaling, Eq.~\eqref{onehalf}, conforms with the Kibble-Zurek mechanism (KZM) that relates the density of topological defects (and, more generally, excitations) to the critical exponents of phase transition and the rate of the quench~\cite{Kibble76,Zurek85}.

The two regimes -- LZ $(\tau_Q\gg\tad$) and KZM $(\tad\ll\tau_Q$) -- switch validity when $\tau_Q\sim\tad$, see Fig.~\ref{fig:LZtoKZM}b. A good indication of the ``border'' between LZ and KZM -- i.e., between adiabatic and non-adiabatic -- is the expected number of excitations: When it is fractional, LZ is a good approximation; When there are several, then KZM should work. 

We expect that, in hardware to implement quantum annealing, one should be able to choose $g$, $J$, $L$, and $\tau_Q$ to cover the range where the ideal quantum Ising chain undergoes a transition from quantum adiabatic LZ behavior (i.e., a successful computation) to non-adiabatic KZM behavior (i.e., a defective computation). Thus, the quench of the Ising chain gives a simple test of adiabatic computing (TAC) for devices that implement quantum annealing. There are other tests that aim at similar goals (e.g., ``quench echo''~\cite{QuanZurek} and the symptoms of entanglement~\cite{Boixo08}). The physically motivated TAC proposed here will be useful in evaluating quantum annealing hardware. 

\section{Results}
\begin{figure*}[t!]
\includegraphics[width=\textwidth]{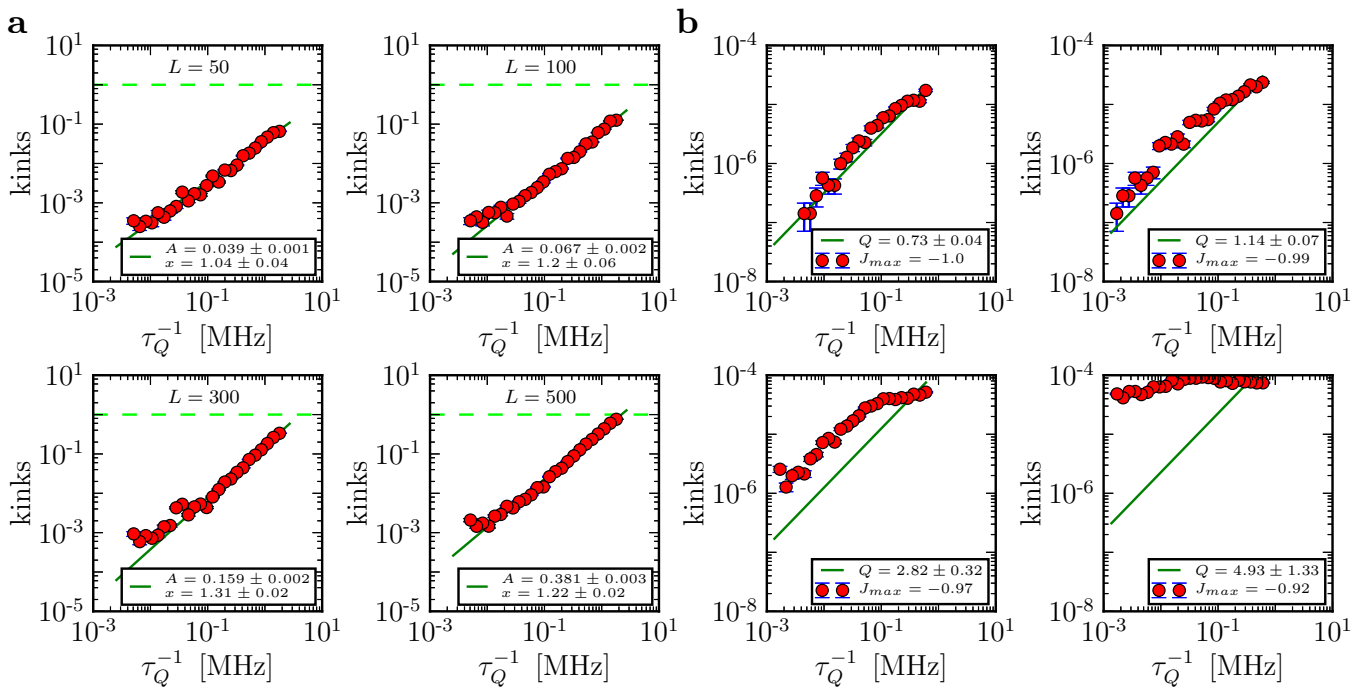}
\caption{
{\bf Defect generation in a quench of the quantum Ising chain on a D-Wave computer.} 
{\bf a.} 
The number of defects versus quench time for two different length chains ($J_{\rm max}=-1$ for all) on the Los Alamos machine. Solid line shows the best fit to the function $A\cdot \tau_Q^{-x}$. 
{\bf b.} 
The number of defects versus quench time for a short chain ($L=4$) for different values of $J_{\rm max}$. Solid line shows the best fit to the function $A\cdot \tau_Q^{-1}$. These results were obtained by averaging over different runs and realizations of the same chain on the chimera graph. Errors are the standard errors of the mean. Note the dramatic change in the behavior between the quenches that start with maximum initial coupling strength (upper right corner) and only 8\% smaller initial coupling strength (lower right corner).
}
\label{fig:results}
\end{figure*}

In D-Wave computers, $L$ can vary from $L=2$ to $L\sim10^3$ and $\tau_Q$ by over two orders of magnitude. Moreover, the maximal value of $J$ at the beginning (and the end) of the quench, respectively, can vary by about two orders of magnitude. We have implemented the quench on both the DW2X-SYS4 (based in Burnaby) and the DW2X (based in Los Alamos), as shown in Fig. \ref{fig:chimera}. The number of kinks in long chains as a function of quench time from the Los Alamos D-Wave DW2X are shown in Fig.~\ref{fig:results}a (see Methods for details and a compilation of results from Burnaby and Los Alamos).

There are several striking and general features of Fig.~\ref{fig:results}. The plots conform well to a power law with the density of kinks proportional to $\tau_Q^{-1}$. This power law differs from the KZM prediction of $\tau_Q^{-1/2}$. Indeed, all of these plots enter the regime where the number of kinks per chain is $\sim0.1$ or less. In this range one expects exponential LZ suppression of excitations, though. We have not found any evidence of such an effect.

Since we do not see the exponential suppression in either open or closed chains of many different lengths from $50$ to $500$ sites, we search for it in very short closed chains, $L=4$, that exhibit LZ crossover for a finite $g$. Additional motivation for this search comes from the scaling $\approx \tau_Q^{-1}$ in Fig.~\ref{fig:results}a. 
It is known~\cite{Avron} that decoherence with energy eigenstates as the pointer states~\cite{PazZurek99} results in a 
$\tau_Q^{-1}$ dependence for the LZ regime. 

The results for $L=4$ closed chains are in Fig.~\ref{fig:results}d. The scaling with $\tau_Q^{-1}$ is still present. This tempts us to regard it as evidence of an anti-crossing in presence of decoherence~\cite{Avron}. The number of kinks, though, seems to be larger than the theory can accommodate. Moreover, we found evidence against this ``LZ with decoherence'' interpretation. For one, quenches with a slightly smaller value of the maximal $J_{\rm max}$ behave differently. The number of kinks can be nearly independent of $\tau_Q$, see Fig.~\ref{fig:results}b. That qualitative change is rather abrupt. Furthermore, quenches with long chains seem to show little dependence on chain length, while one expects the kink number to increase with chain length. 

We do not see how these features can be accommodated within any known general theories (e.g., LZ, KZM, LZ with decoherence). Furthermore, we find significant differences between Ising chains of the same length implemented using different ``spins'' (i.e., Josephson junctions) on the D-Wave chip, as well as differences between the Los Alamos and Burnaby machines. In particular, the number of defects, as well as the scatter, is significantly smaller in the Los Alamos machine compared to the Burnaby machine for similar Ising chains, quench rates, etc. (see Methods).

\section{Quantum Ising chain in a hostile environment}
Many factors can be contributing to this unusual behavior: Heating, randomness in couplings, eigenstate decoherence, local decoherence, self-interactions, non-Markovian effects, noise, etc. Many of these issues will likely be encountered in other settings. We note that, in our case, some of them can be ruled out, while others can not. The following discussion is inspired by our thinking of what can happen to a quantum Ising chain implemented on a D-Wave chip. Essentially, we discuss the behavior of Ising chains that are not completely isolated from their environment. We do not aim to be exhaustive: We have selected models of decoherence that can be described relatively simply (which does not mean that they can be readily solved!). We have also focused on models that can be simply parametrized (thus, for example, we have avoided discussing ``mixtures'' of models that -- like models of noise -- have several components).

This selection of what is to be discussed is in accord with the goal we have -- understanding of the role of external factors in the dynamics of phase transitions as represented by a quantum Ising chain. This may come handy not just in benchmarking of adiabatic quantum computers, but also in future condensed matter experiments where quantum many-body systems are driven through a symmetry breaking transitions in presence of the inevitable coupling with their environment. Thus, while the D-Wave chip is ``on  our mind'', we feel that many of the problems we shall encounter in the discussion of its physics will be also encountered in other settings.

\begin{figure}
\vspace{-0cm}
\includegraphics[width=\columnwidth,clip=true]{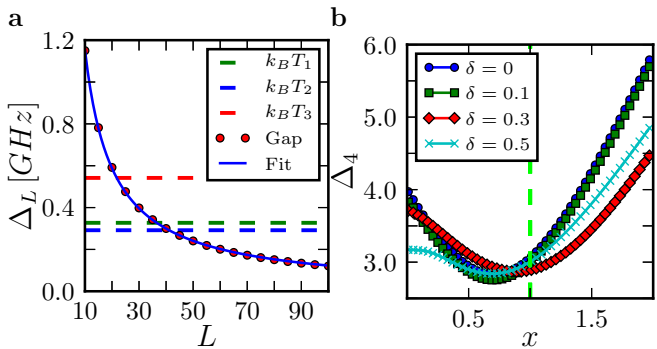}
\vspace{-0cm}
\caption{
{\bf Energy gaps.}
{\bf a.} 
The gap $\Delta_L$ at the critical point $s_c$ for the Ising chain implemented on the D-Wave chip.
Red dots where obtained from numerical calculations using the D-Wave protocol, Fig.~1(a). The solid line is the fit $\Delta_L=\Delta_0\cdot L^{-x}$, where $\Delta_0=(10.84\pm 0.06)\,$[GHz] and $x=0.973\pm 0.002$. Dotted lines show the thermal energy for the DW2X (Los Alamos) chip: $T_1=(15.7\pm1.0)\,\text{mK}$, DW2X (Los Alamos) chip: $T_2=(14\pm1.15)\,\text{mK}$ and DW2X-SYS4 (Burnaby): $T_3=(26 \pm 5)\, \text{mK}$. 
{\bf b.} 
The gap $\Delta_4$ for the closed random model~(\ref{gHising}) where both 
$g_i\in\left[x-\delta,x+\delta\right]$ and $J_i\in\left[1-\delta,1+\delta\right]$ are drawn from a uniform distribution. The disorder has weak perturbative effect even at the anti-crossing center.
}
\label{fig:heating}
\end{figure}

{\bf Thermal excitation.~}
Heating of the Ising chains is an obvious culprit that would add excitations -- generate kinks. We do not believe that, in the D-Wave setting, it is dominant. The heating will be most effective near the critical point, as the temperatures of the two D-Wave chips we have worked with exceed the size of the gap only in its vicinity for the chains we studied: Figure~\ref{fig:heating} shows the minimal energy gap (near the quantum critical point) for different lengths of the quantum Ising chain. Thus, kink generating transitions will be only effective for a period of time that is roughly proportional to $\tau_Q$. If this effect was dominant it should result in the number of kinks increasing with $\tau_Q$. We observe the opposite trend (e.g., $~\tau_Q^{-1}$ in the Los Alamos chip). Furthermore, for very short chains (e.g., squares) there is over an order of magnitude difference between the minimal gap and $k_BT$ of the chip, suppressing thermal excitations.

For above reasons, we conclude that ``heating'' is unlikely to be the dominant effect behind the generation of kinks above the Landau-Zener theory predictions.

{\bf Coupling to the spins not in the chain.~}
It is known that the spins on the D-Wave chip also couple to the spins from which they are nominally decoupled. That is, setting the coupling $J_{kl}=0$ between spins $k$ and $l$ does not guarantee that this coupling is indeed negligible. There are also reasons to believe that this coupling is predominantly ``Ising'' ($\sim \sigma^z_k \sigma^z_l$) rather than, e.g., Heisenberg.

We believe we have seen evidence of such spurious couplings in the behavior of the Ising chains. For instance, the ``compact chains'' (that cover relatively small area of the chip) yield fewer kinks than ``spread out chains'' of the same length. This would happen if the spurious coupling with spins that should be decoupled from the chain resulted in the couplings between different fragments of the chain. This would have two related effects: The Hamiltonian of Eq.~(2) is no longer the whole story (as it will be dressed with the couplings to the spins from which it should be nominally decoupled). We will not model this effect (in part because it requires a detailed account of how these spurious couplings occur and, in part, we believe it may turn out to be too D-Wave-specific).

The second effect that we will model recognizes that such ``ghost spins'' act as an environment that will decohere fragments of the quantum Ising chain -- ``ghost spins'' monitor the orientation of the spins inside the chain. This is of interest, and is likely to be ubiquitous in other realizations of the quantum Ising systems, both in condensed matter and quantum information processing devices.

\begin{figure}
\vspace{-0cm}
\includegraphics[width=\columnwidth]{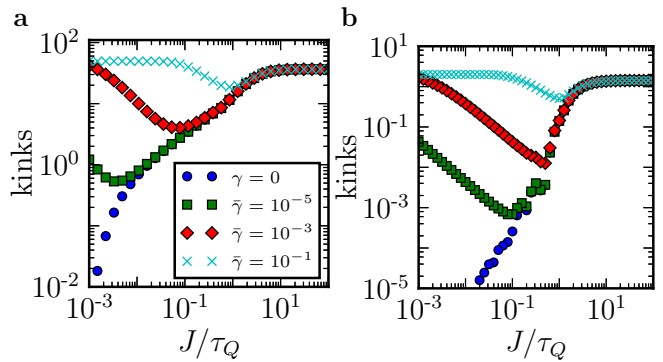}
\vspace{-0cm}
\caption{
{\bf Ghost spins - defects with decoherence.}
{\bf a.} 
The number of kinks as a function of quench rate $J/\tau_Q$ for different decoherence strengths $\gamma$ and $L=100$. Here, $g(t)=J(1-t/\tau_Q)$ and $\bar{\gamma}=\gamma/J$.
{\bf b.} 
The same as in {\bf a.} but for a periodic chain of length $L=4$ (i.e., a ``square'').
}
\label{fig:decoherence}
\end{figure}

We model this effect in Methods for both open chains of varying length, see Fig.~\ref{fig:decoherence}(a), and closed ``squares'', see Fig.~\ref{fig:decoherence}(b). There is a generic pattern that emerges: When decoherence due to ``ghost spins'' acts for sufficiently long time, the number of kinks begins to increase with $\tau_Q$ until it saturates at $\simeq L/2$. A similar effect was studied before in Ref.~\cite{Adolfo16} where it was described as anti-Kibble-Zurek behaviour.

{\bf Randomness in the Hamiltonian.~}
It is now known that the implementation of the Ising Hamiltonian, Eq.~(2), suffers from errors both in the value of the couplings between spins [i.e., $J(t)$] and the bias field $g(t)$. These errors are difficult to characterize in detail especially in the critical region where $g(t)\approx J(t)$. They tend to be several percent of the maximal values of $g$ and $J$~\footnote{Private communication with D-Wave Inc.}. The relative error, though, in $g-J$ near the critical point, however, could be large. 

Such randomness has a profound effect on the dynamics and kink generation that to some extent has been analyzed~\cite{JDRandom,FazioRandom,JDDec}. Random couplings and transverse fields, which we allow for in the Hamiltonian of Eq.~\eqref{gHising}, alters the universality class for a long enough chain. We note that we take the randomness in $g_i(t)$ and $J_i(t)$ to include static random fluctuations around uniform $g(t)$ and $J(t)$, respectively. The number of kinks after a quench is no longer a power law $\sim\tau_Q^{-1/2}$ predicted by KZM for a homogeneous chain but a logarithmic decay 
$\sim(\ln\tau_Q)^{-2}$~\cite{JDRandom,FazioRandom,JDDec}. 

This slow decay might possibly explain the absence of the exponential LZ decay for long enough chains: In the presence of disorder, the adiabaticity estimate in Eq. (6) is no longer valid and much longer quench times are required. However, the longer chains seem to conform to a power law rather than the logarithmic decay and, what is more important, the power law persists even in short chains like the $L=4$ squares. As seen in Fig.~\ref{fig:heating}(a), the square has a relatively large gap even at the anti-crossing so the disorder could only have a weak perturbative effect on the outcome of the quench, see Fig.~\ref{fig:heating}(b). Thus, we conclude that disorder is not the main culprit for the observed discrepancies with respect to the pure Ising chain.

{\bf Decoherence in energy eigenstates.~}
A model of an anti-crossing with decoherence via Lindblad superoperators that are diagonal in the instantaneous energy eigenstates turns out to be exactly solvable~\cite{Avron}. Moreover, for short chains (i.e., squares), decoherence that favors energy eigenstates can be relevant (as it tends to set in whenever the separation of energy levels is large compared to the other relevant energy scales~\cite{PazZurek99}).

In this regime, the probability of a transition to the excited state is given by the equation
\be 
p=\frac{\varepsilon \hbar}{2\Delta_L^2}~Q\left(\frac{\hbar\gamma}{\Delta_L}\right),
\label{avron}
\ee
where $\gamma$ and $\varepsilon\sim\tau_Q^{-1}$ are the decoherence and transition rates, respectively, and $Q$ is a simple function with a maximum value of $\sim0.65$~\cite{Avron}. Our results with squares yield values of $Q$ that are close to $Q\sim 1$ and that sometimes ``dip'' to within the region below $0.65$ consistent with the equation above. We note that our estimates of the parameters in Eq.~(\ref{avron}) can be significantly affected by the caveats listed above, so we cannot rule out significance of this model for squares. 

In particular, the probability of kink formation for both ferro and anti-ferro cases exhibits the same quench-rate dependence ($\tau_Q^{-1}$) consistent with Eq.~(\ref{avron}) only in the Los Alamos machine and when the scale of $J$ is set to its maximal range, see Fig.~\ref{fig:diff}(b). However, even a relatively modest change of that scale from the maximum leads to a fairly dramatic change in the behavior undermining hope in the utility of Eq.~(\ref{avron}) for the problem at hand, see Fig.~3(b).

\jd{

}

\section{Discussion}
Complex behavior of quantum annealers demands {\em global} tests of adiabaticity and quantumness, as even when components of the device work, their integration raises questions of decoherence, control, and what ``slow enough'' is. We propose a global test based on a quench in the quantum Ising chain. It can assess reliability of the whole device. Such general tests will prove valuable in establishing adiabaticity and benchmarking$/$comparing different implementations of adiabatic quantum computers expected in the near future. 

In spite of the outcome of the TAC, D-Wave may, in some cases, find the right or at least approximate, solutions to problems. Obviously, a more precise implementation would result in a more successful adiabatic quantum computation$/$quantum anneal. Indeed, the noticeable decrease in the number of defects between the tests of Burnaby and Los Alamos machines is likely due to the improvements in hardware. One can hope that the next generation of quantum annealers will be even better. \\

\vspace{2mm}

{\noindent \bf Methods}\\

{\bf Numerical simulations.~}
To obtain the results in Fig.~\ref{fig:LZtoKZM}b, we first brought the Hamiltonian~(\ref{Hising}) into its 
fermionic representation~\citep{AdvPhys}, 
\be
\begin{split}
	H &= 2 \sum_{n=1}^L g_n c_n^\dag c_n -\sum_{n=1}^Lg_n\\
	&- \sum_{n=1}^{L-1} J_n 
	\left(
	c_n^\dag c_{n+1} + c_{n}^\dag c_{n+1}^\dag + \text{h.c.}
	\right),
\end{split}
\ee
using the Jordan-Wigner transformation~\cite{zwolak12}
\bea 
\sigma^z_n &=&
\left(c_n^\dag+c_n\right)
\prod_{m<n}
\left(1-2c_m^\dag c_m\right),\\
\sigma^x_n &=&
1-2c_n^\dag c_n,
\label{sigmazJW}
\eea
where $c_n$ ($c_n^{\dagger}$) is the fermionic annihilation (creation) operator for site $n$. For the quadratic correlation functions $x_{pq}:=\langle c_p^\dag c_q \rangle$ and $y_{pq}:=\langle c_p^\dag c_q^\dag \rangle$, this gives the closed system of equations
\be
\begin{split}
	\label{x}
	i\dot{x}_{p,q} &= -J_{p}\,x_{p+1,q} - J_{p-1}\,x_{p-1,q} + J_q\,x_{p,q+1}  \\  
	&+ J_{q-1}\,x_{p,q-1} + J_q\,y_{p,q+1} - J_{q-1}\,y_{p,q-1} \\
	&+ J_p \,y_{p+1,q}^* - J_{p-1}\, y_{p-1,q}^* \\ 
	&+ 2\left(g_p-g_q\right)x_{p,q}, \quad q\geq p;
\end{split}
\ee
and
\be
\begin{split}
	\label{y}
	i\dot{y}_{p,q} &= -J_p\,y_{p+1,q} - J_{p-1}\,y_{p-1,q} - J_q\,y_{p,q+1}  \\
	&- J_{q-1}\,y_{p,q-1} - J_q\,x_{p,q+1} + J_{q-1}\, x_{p,q-1} \\
	&+ J_p\,x_{p+1,q}^* - J_{p-1} x_{p-1,q}^* - J_p\,\delta_{p+1,q} \\
	&+ 2(g_p+g_q)y_{p,q}, \quad q>p ,
\end{split}	
\ee
with $y_{pp}=0$. 
The above equations are solved with the initial condition corresponding to the system's ground
state when $J>0$ and with the boundary conditions $c_0=c_{L+1}=0$. To carry out numerical computations, 
we used an adaptive Adams method from LSODA. 
Finally, the number of kinks was obtained from 
\be
\text{kinks} = \frac{L-1}{2}-\sum_{p=1}^{L-1}\Re\left(x_{p,p+1}+y_{p,p+1}\right).
\ee 
Here $\Re$ means real part.
Both the ground state and the gap depicted in Fig.~\ref{fig:LZtoKZM}(a) where calculated 
using techniques described in Ref.~\cite{lieb61}. 

Thus, to compute the number of kinks we used the following formula

\begin{equation}
\label{kinksT}
\text{kinks}=\frac{N|J_{\text{max}}|+E}{2|J_{\text{max}}|},
\end{equation}
$N$ is the numbers of couplings in the chain. 
The final energy $E$ can be read in \emph{directly} from the D-Wave solver.

\begin{figure}[tp!]
	\vspace{-0cm}
	\includegraphics[width=\columnwidth,clip=true]{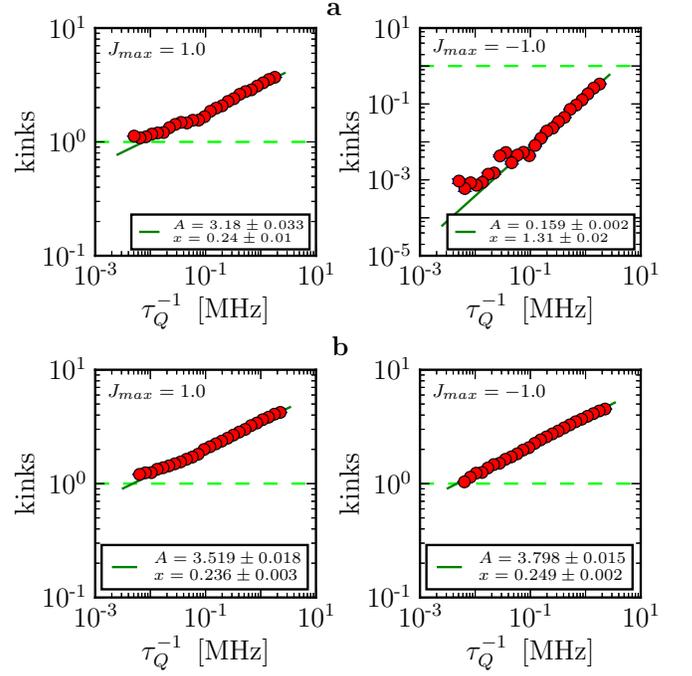}
	\vspace{-0cm}
	\caption{
	{\bf Comparison between the same D-Wave architectures ($\mathbf{L=300}$).}
	{\bf a.} DW2X system based in Los Alamos. The number of kinks is different
	for ferro and anti-ferro cases. Smaller number of defects when $J<0$ suggest
	that the Los Alamos chips performs better in this regime.
	{\bf b.} DW2X system based in Burnaby. As one would expect, the number of 
	kinks is roughly the same for both ferro and anti-ferro cases. 
	}
	\label{fig:diff}
\end{figure}

{\bf Burnaby versus Los Alamos chip.~} 
One would expect different chips of the same generation of annealers to generate roughly
the same number of kinks for ferro and anti-ferro cases. However, the DW2X based in Los Alamos seems to perform better (i.e., generates less kinks) when $J<0$, see Fig.~\ref{fig:diff}(a). The Burnaby machine, though, has the same behavior for ferro and anti-ferro cases, see Fig.~\ref{fig:diff}(b).

\begin{figure}[tp!]
	\vspace{-0cm}
	\includegraphics[width=\columnwidth,clip=true]{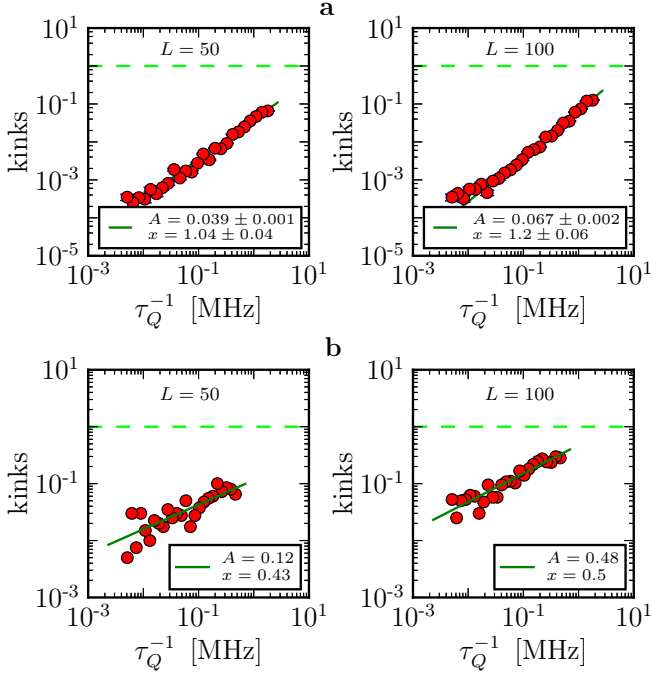}
	\vspace{-0cm}
	\caption{
		{\bf Comparison between different D-Wave architectures ($\mathbf{J_{\text{max}}=-1}$).}
		{\bf a.} DW2X system based in Los Alamos.
		{\bf b.} A previous generation DW2X-SYS4 in Burnaby.
		 The DW2X chip is better, i.e. it produces less kinks. Moreover, these two architectures exhibit
		 different quench-time dependence ($\tau_Q^{-1}$ versus $\tau_Q^{-1/2}$).	 
	}
	\label{fig:diff2}
\end{figure}

Thus, chips that belong to different architectures may behave differently. 
For instance, Fig.~\ref{fig:diff2} compares the DW2X based in Los Alamos and a previous
generation DW2X-SYS4 in Burnaby. Not only do the number of kinks differ between these
two systems but it also exhibits different quench-time dependence ($\tau_Q^{-1}$ versus $\tau_Q^{-1/2}$).
%
%
%


{\bf Decoherence by ``ghost spins''.~} 
Numerical results presented in Fig.~\ref{fig:decoherence} are obtained using the following Linbdlad master equation~\cite{zwolak12}:
\begin{equation}
\label{L2}
	\dot{\rho}(t) = 
	\frac{1}{i\hbar}\left[H(t), \rho(t)\right] + 
	\gamma D[\rho(t)], 
\end{equation}
where the superoperator is
\begin{equation}
\label{diss2}
	D[\rho(t)] = -\frac{1}{2}\sum_{n=1}^L\left[\sigma_n^z,\left[\sigma_n^z,\rho(t)\right]\right]
\end{equation}
and $H(t)$ takes the form
\be 
H(t)=-\sum_{n=1}^{L}g_n(t)\sigma^x_n-\sum_{i=n}^{L-1}J_n(t)\sigma^z_n\sigma^z_{n+1},
\label{gHising}
\ee
where we allow for time and spatial dependence in both $J_n$ and $g_n$.

Expectation values, $\langle O\rangle={\rm Tr}(O\rho)$, of an operator $O$ evolve according to 
\be 
\frac{d}{dt}\langle O\rangle=
\frac{1}{i\hbar} \langle [O,H] \rangle -
\frac{\gamma}{2}\sum_{n=1}^L
\langle  [\sigma^z_n,[\sigma^z_n,O]] \rangle.
\label{dOdt}
\ee
This equation is solved using the Jordan-Wigner transformation~\cite{lieb61}
\bea 
\sigma^z_n &=&
\left(c_n^\dag+c_n\right)
\prod_{m<n}
\left(1-2c_m^\dag c_m\right),\\
\sigma^x_n &=&
1-2c_n^\dag c_n,
\label{sigmazJW}
\eea
where $c_n$ ($c_n^{\dagger}$) is a fermionic annihilation (creation) operator. For an open chain, the above transformation brings the Hamiltonian, Eq.~\eqref{gHising}, to the following form~\citep{AdvPhys}
\be
 \begin{split}
	H &= 2 \sum_{n=1}^L g_n c_n^\dag c_n -g_n L \\
	  &- \sum_{n=1}^{L-1} J_n 
      \left(
		c_n^\dag c_{n+1} + c_{n}^\dag c_{n+1}^\dag + \text{h.c.}
      \right).
 \end{split}
\ee
The string operators in Eq.~(\ref{sigmazJW}) cancel out in the Lindblad contribution to
the right hand side of Eq.~(\ref{dOdt}), hence the quadratic fermionic correlation functions
$x_{pq}:=\langle c_p^\dag c_q \rangle$ and $y_{pq}:=\langle c_p^\dag c_q^\dag \rangle$
satisfy a closed set of equations ($q\geq p$), 
\be
 \begin{split}
 	\label{x}
	i\dot{x}_{p,q} &= -J_{p}\,x_{p+1,q} - J_{p-1}\,x_{p-1,q} + J_q\,x_{p,q+1}  \\  
				   &+ J_{q-1}\,x_{p,q-1} + J_q\,y_{p,q+1} - J_{q-1}\,y_{p,q-1} \\
				   &+ J_p \,y_{p+1,q}^* - J_{p-1}\, y_{p-1,q}^* \\ 
				   &+ 2\left(h_p-h_q\right)x_{p,q} + \gamma \mathcal{D}[x_{pq}],
 \end{split}
\ee
where the Lindblad superoperator $\mathcal{D}[x_{pq}]$ reads
\begin{equation}
	\mathcal{D}[x_{pq}]
	=
	\begin{cases}
	1-2x_{pp} \quad & \text{if } \quad p=q,  \\
	2\Re(y_{pq})-2|q-p|x_{pq}	& \text{if } \quad  q>p,
	\end{cases}
\end{equation}
together with $y_{pp}=0$ and ($q>p$)
\be
  \begin{split}
  	\label{y}
	i\dot{y}_{p,q} &= -J_p\,y_{p+1,q} - J_{p-1}\,y_{p-1,q} - J_q\,y_{p,q+1}  \\
	&- J_{q-1}\,y_{p,q-1} - J_q\,x_{p,q+1} + J_{q-1}\, x_{p,q-1} \\
	&+ J_p\,x_{p+1,q}^* - J_{p-1} x_{p-1,q}^* - J_p\,\delta_{p+1,q} \\
	&+ 2(h_p+h_q)y_{p,q} + \gamma \mathcal{D}[y_{pq}], 
  \end{split}	
\ee
with  $\mathcal{D}[y_{pq}]=2\Re(x_{pq})-2|q-p|y_{pq}$. 

These equations are to be solved with the initial condition corresponding to the system's ground
state when $J>0$ and with the boundary conditions $c_0=c_{L+1}=0$~\cite{lieb61}. 
The number of kinks is then given by
\be
 \text{kinks} = \frac{L-1}{2}-\sum_{p=1}^{L-1}\Re\left(x_{p,p+1}+y_{p,p+1}\right).
\ee 

\noindent{\bf Bibliography}

%
%
%
%
{\noindent \bf Acknowledgments}\\
We appreciate fruitful discussions with our Los Alamos colleagues Marcus Daniels, Seth Lloyd, Scott Pakin, and Rolando Somma, as well as Edward Dahl and Trevor Lanting of D-Wave Systems. Work of J.D. and B.G. was supported by Narodowe Centrum Nauki (National 
Science Center) under Project No. 2016/23/B/ST3/00830 and 2016/20/S/ST2/00152, respectively. Work of W.H.Z. was supported by the US Department of Energy under the Los Alamos LDRD program. This research was supported in part by PL-Grid Infrastructure. \\

{\noindent \bf Author contributions}\\
B.G., J.D., W.H.Z., M.Z. contributed equally to this work.\\

{\noindent \bf Additional information}\\

{\noindent \bf Competing interests:} The authors declare no competing interests.

\end{document}